\documentclass{article}
\usepackage{spconf,amsmath,graphicx,hyperref}
\usepackage{xcolor}
\usepackage{cite}
\usepackage{amssymb}
\usepackage{algorithm}
\usepackage{xcolor}
\usepackage{booktabs}
\usepackage{url}
\usepackage{pgfplots}
\pgfplotsset{compat=1.18}  
\usepgfplotslibrary{fillbetween}
\usepackage{booktabs}  
\usepackage{algorithm}
\usepackage{algpseudocode}
\usepackage{enumitem}
\usepackage{pifont} 

\title{Few-Shot and Pseudo-Label Guided Speech Quality Evaluation with Large Language Models}
\name{%
\begin{tabular}{c}
\it
Ryandhimas E. Zezario$^{\dagger}$,  Dyah A. M. G. Wisnu$^{\dagger,\ast}$, Szu-Wei Fu$^{\ddagger}$,\\ Sabato Marco Siniscalchi$^{\S}$
, Hsin-Min Wang$^{\dagger}$, Yu Tsao$^{\dagger}$
\end{tabular}
}

\address{
    $^{\dagger}$Academia Sinica,   $^{\ast}$National Chengchi University,
    $^{\ddagger}$NVIDIA,    
    $^{\S}$University of Palermo
    \\
}
\begin{document}
%
\maketitle
\begin{abstract}
In this paper, we introduce GatherMOS, a novel framework that leverages large language models (LLM) as meta-evaluators to aggregate diverse signals into quality predictions. GatherMOS integrates lightweight acoustic descriptors with pseudo-labels from DNSMOS and VQScore, enabling the LLM to reason over heterogeneous inputs and infer perceptual mean opinion scores (MOS). We further explore both zero-shot and few-shot in-context learning setups, showing that zero-shot GatherMOS maintains stable performance across diverse conditions, while few-shot guidance yields large gains when support samples match the test conditions. Experiments on the VoiceBank-DEMAND dataset demonstrate that GatherMOS consistently outperforms DNSMOS, VQScore, naive score averaging, and even learning-based models such as CNN-BLSTM and MOS-SSL when trained under limited labeled-data conditions. These results highlight the potential of LLM-based aggregation as a practical strategy for non-intrusive speech quality evaluation.

\end{abstract}

\section{Introduction}
\label{sec:intro}
Speech quality has long been an important indicator for evaluating various speech processing applications, including speech enhancement \cite{loizou2013speech}, hearing aid (HA) devices \cite{kates2014hearingb}, and telecommunications \cite{polqa_2013}. While human-based evaluation remains the gold standard, it requires a sufficient number of listeners to obtain reliable and generalized scores. With the increasing availability of large-scale audio datasets with corresponding human-labeled scores, deep learning models have attracted significant attention for deploying non-intrusive speech quality prediction systems \cite{ref_51, mosa-net, fu2024selfsupervised, ssl-mos, 10890590, chen2025audio}. More recently, large-scale pre-trained audio models such as wav2vec 2.0 \cite{wav2vec} and Whisper \cite{Whisper} have emerged as powerful acoustic feature extractors, further boosting the performance of non-intrusive quality assessment models.  Nevertheless, having a sufficient number of training samples remains crucial for achieving robust and reliable model performance.


Recently, large language models (LLM) such as GPT-4o and GPT-5 have demonstrated strong reasoning capabilities and the ability to integrate multimodal information. Several works have begun exploring their use in speech-related tasks, including pathological speech detection and zero-shot intelligibility assessment \cite{amiri2025exploringincontextlearningcapabilities, 10889809}. For example, \cite{10889809} converted audio into text-based representations and prompted ChatGPT to judge naturalness. While promising, such studies rely on direct prompting and lack explicit mechanisms to incorporate acoustic features or auxiliary predictors, which limits their robustness. Moreover, our preliminary experiments indicate that ChatGPT struggles with raw audio, requiring intermediate features as proxies. However, handcrafted features (e.g., root mean square (RMS), zero-crossing rate (ZCR), and mel-frequency cepstral coefficients (MFCC)) only provide coarse summaries of the signal and often correlate weakly with perceptual judgments. This motivates the need for a framework that can integrate richer acoustic descriptors with complementary predictors, allowing LLMs to make more reliable and fine-grained quality estimations.

\graphicspath{ {./images/} }
\begin{figure*}[t]
\centering
\includegraphics[width=15cm]{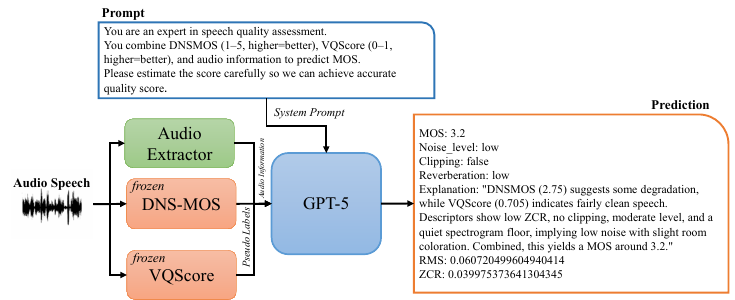} 
\caption{Detailed architecture of GatherMOS} 
\label{fig:cross-domain}
\end{figure*}

Given these constraints, we aim to further investigate how LLM can be leveraged for speech quality estimation. In particular, we propose GatherMOS, a framework that acts as a meta-evaluator by aggregating diverse weak signals into stronger and more reliable predictions. GatherMOS combines acoustic feature representations with pseudo-labels obtained from DNSMOS \cite{9414878} and VQScore \cite{fu2024selfsupervised}, enabling the LLM to reason over structured inputs rather than relying solely on coarse audio information or text proxies. We specifically choose DNSMOS and VQScore for their lightweight design and independence from large-scale models, which enhances practicality. By leveraging LLM’s reasoning capabilities, GatherMOS integrates information from both descriptors and pseudo-labels to refine MOS estimation. Additionally, we explore whether incorporating more detailed acoustic features, such as MFCC and spectrogram images, along with few-shot in-context information, can further improve performance. Experimental results confirm that GatherMOS outperforms DNSMOS, VQScore, and naive score averaging, as well as learning-based models trained on limited data. We also observe that mismatched few-shot samples can reduce prediction accuracy, whereas richer acoustic features consistently improve results. These findings suggest that GatherMOS has potential as an LLM-based meta-evaluator for non-intrusive speech quality, complementing existing methods and pointing toward more scalable evaluation with reduced dependence on annotated data.

The remainder of this paper is organized as follows. Section II presents the proposed GatherMOS method. Section III describes the experimental setup and results. Finally, Section IV provides conclusions and future work.

\section{GatherMOS}
\subsection{Zero-shot GatherMOS}
Given an input waveform $x \in \mathbb{R}^T$, we extract several acoustic descriptors that summarize temporal, spectral, and perceptual information. The RMS reflects the overall signal energy, while the ZCR indicates the level of noisiness or voicing. Duration and clipping detection are included to capture temporal length and signal distortion when the maximum amplitude approaches unity. In addition, 13 MFCC are computed and averaged across frames to represent spectral envelope characteristics. Finally, mel-spectrogram statistics, including per-bin mean and variance as well as global maximum and minimum values, are extracted to capture spectral distribution patterns. In addition to handcrafted descriptors, we incorporate two lightweight predictors DNSMOS ($s_{\text{DNS}} \in [1,5]$) and VQScore($s_{\text{VQ}} \in [0,1]$). These pseudo-labels act as weak supervision signals that correlate with perceptual quality while remaining computationally efficient.  

We aggregate descriptors and pseudo-labels using an LLM, specifically GPT-5. The inputs are serialized into a structured textual prompt, and the model is instructed to output a continuous MOS prediction $\hat{s}$ along with auxiliary explanatory attributes as shown in Fig. 1. Formally, GatherMOS is expressed as:
\begin{equation}
    \hat{s}, a = f_{\text{LLM}}\!\left( \{d_i\}_{i=1}^N, \; s_{\text{DNS}}, \; s_{\text{VQ}} \right),
\end{equation}
where $\{d_i\}_{i=1}^N$ are extracted descriptors, $s_{\text{DNS}}$ and $s_{\text{VQ}}$ denote pseudo-labels, and $a$ represents additional attributes (e.g., noise level, clipping, reverberation).  

\subsection{Few-Shot GatherMOS}
When representative few-shot examples ${(x_j, s_j)}$ are available, they are incorporated into the prompt to provide in-context guidance during inference. Specifically, these few-shot samples are provided as input to the LLM along with the other descriptors:
\begin{equation}
    \hat{s}, a = f_{\text{LLM}}\!\left( \{d_i\}_{i=1}^N, \; s_{\text{DNS}}, \; s_{\text{VQ}}, \; \{(x_j, s_j)\}_{j=1}^K \right),
\end{equation}
where $\{(x_j, s_j)\}_{j=1}^K$ denotes the few-shot support set consisting of $K$ labeled examples. In this formulation, the support set is provided solely as part of the model input context and does not involve any update to model parameters. The LLM is therefore guided by both pseudo-labels and few-shot examples at inference time.

\section{Experiments}
\label{sec:exp}
\subsection{Experimental setup}
The proposed approaches are evaluated on the VoiceBank-DEMAND dataset \cite{Voicebank_Demand}, which is also included in the test evaluation of the VoiceMOS Challenge 2024 \cite{10832295}. We select a test set of 200 utterances, consisting of clean speech, noisy speech corrupted by four noise types (car, white, music, and cafeteria) at an SNR of 0 dB, and enhanced speech produced by five representative enhancement systems: BSSE \cite{hung2022boosting}, MPSENet \cite{lu2023mp}, CMGAN \cite{cao2022cmgan}, DEMUCS \cite{defossez2020real}, and Wiener filtering \cite{loizou2013speech}. Human evaluation is conducted with ten listeners (six male and four female), where each utterance is rated by five listeners. For objective assessment, we report two correlation-based metrics: the linear correlation coefficient (LCC) and Spearman’s rank correlation coefficient (SRCC) \cite{srcc}, where higher values indicate stronger alignment between predicted scores and ground-truth human ratings.
\subsection{Few samples evaluation}

In this experiment, we first use a small set of ten speech samples covering different acoustic environments and enhancement systems. Starting with a limited number of samples enables a controlled comparison with the baselines, facilitates detailed inspection of individual predictions, and allows us to verify the behavior of GatherMOS before proceeding to full-scale evaluation. For DNSMOS \cite{9414878}, we adopt the publicly available checkpoint provided on GitHub to obtain evaluation scores. Similarly, for VQScore \cite{fu2024selfsupervised}, we use the official GitHub checkpoint with its recommended configuration. For NaiveEnsemble, we compute the average of the scores predicted by DNSMOS and VQScore. To ensure fairness, we normalize the outputs of each model so that both scores are on the same scale prior to averaging. For the proposed approach, we investigate two variants of GatherMOS. GatherMOS-ZS refers to the zero-shot setting that leverages basic acoustic features, including RMS, ZCR, clipping ratio, and duration. GatherMOS-FS corresponds to the few-shot setting, which incorporates the same acoustic features as GatherMOS-ZS together with a small support set of labeled examples. Specifically, three representative few-shot examples corresponding to low, medium, and high MOS levels are used. As illustrated in Fig. 2, these examples are selected to span different quality ranges. The few-shot examples are unseen from the test set to mimic a more realistic scenario, and they are selected from the UDASE task of the 7th CHiME Challenge \cite{chime7-udase, chime7-evaluation}. 

\begin{table}[t]
\caption{Evaluation scores of GatherMOS with a few sample test set.}
\begin{center}
 \begin{tabular}{ccc} 
 \hline
 \textbf{Systems}& \textbf{LCC} & \textbf{SRCC}  \\ [0.5ex] 
 \hline
DNSMOS &0.5538&0.5231\\ 
VQScore &0.4631&0.6359\\ 
NaiveEnsemble & 0.6255	&0.5490\\
GatherMOS-ZS&0.6310&0.6420\\
GatherMOS-FS&\textbf{0.6653}&\textbf{0.8473}
\\ \hline
\end{tabular}
\end{center}
\end{table}

Table 1 summarizes the evaluation results in terms of SRCC and LCC. The baselines, DNSMOS and VQScore, show moderate correlation performance. NaiveEnsemble, which simply averages the outputs of the two baselines, provides a marginal improvement. In addition, GatherMOS-ZS outperforms all baselines, showing the benefit of incorporating acoustic features together with pseudo-labels into the LLM framework. Furthermore, using a small set of few-shot examples further improves prediction accuracy. Specifically, GatherMOS-FS increases the SRCC substantially, reaching 0.8473, which highlights the effectiveness of few-shot examples in guiding the LLM toward more accurate speech quality prediction.

\begin{table}[t]
\caption{Evaluation scores of GatherMOS with all sample test set.}
\begin{center}
 \begin{tabular}{ccc} 
 \hline
 \textbf{Systems}& \textbf{LCC} & \textbf{SRCC}  \\ [0.5ex] 
 \hline
DNSMOS&0.6021&0.5314\\ 
VQScore&0.5753&0.4476\\ 
NaiveEnsemble&0.6106&0.5177\\ 
CNN-BLSTM&0.3192&0.2971\\
MOS-SSL&0.4888&0.4732\\
GatherMOS-ZS&0.6439&0.6014\\ 
GatherMOS-ZS*&\textbf{0.6495}&\textbf{0.6069}\\ 
GatherMOS-FS&0.5653&0.4770
\\ \hline
\end{tabular}
\end{center}
\end{table}

\graphicspath{ {./images/} }
\begin{figure}[t]
\centering
\includegraphics[width=6.25cm]{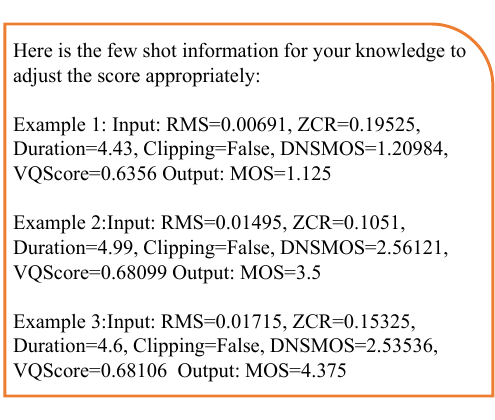} 
\caption{Few-shot prompt examples used for in-context guidance.} 
\label{fig:cross-domain}
\end{figure}

\subsection{All samples evaluation}
In this experiment, we evaluate GatherMOS and baseline methods on the full set of speech samples. We also introduce GatherMOS-ZS*, which incorporates additional MFCC and spectrogram features into the model. To ensure prediction stability, test samples are evaluated in minibatches of ten, with the LLM session reset between minibatches, which helps mitigate potential cross-sample conditioning effects in LLM inference. In addition, we build two trainable neural speech assessment baselines: a CNN-BLSTM-ATT model following the architecture in \cite{ref_51}, and a MOS-SSL model following \cite{ssl-mos}, which integrates a wav2vec module. For these two trainable models, we investigate whether it is feasible to train them from scratch under very limited labeled-data. Specifically, we train CNN-BLSTM and MOS-SSL using the same three labeled samples as a stress-test to evaluate their feasibility under extreme low-resource conditions, rather than as a competitive fully supervised setting.

Table 2 reports the performance in terms of SRCC and LCC. DNSMOS and VQScore achieve moderate correlation, and NaiveEnsemble provides marginal improvements. CNN-BLSTM and MOS-SSL, trained with limited data, show weaker correlation compared to other methods. GatherMOS-ZS, which uses basic acoustic features (RMS, ZCR, clipping ratio, and duration), consistently outperforms all baselines. GatherMOS-ZS*, which further includes MFCC and spectrogram features, provides a slight improvement over GatherMOS-ZS, showing the contribution of richer acoustic representations. Next, GatherMOS-FS, which leverages a small few-shot support set, performs worse than the zero-shot variants on the full test set. As shown in Table 1, GatherMOS-FS achieves the best performance on the few-shot subset, indicating that the model can exploit few-shot examples effectively when the support set closely matches the test samples. The drop in performance on the full set suggests that the few-shot examples do not sufficiently represent the broader distribution, leading to domain bias and overfitting to the support examples. This highlights a tradeoff between in-domain few-shot adaptation and cross-condition generalization. In contrast, the zero-shot variants, especially GatherMOS-ZS*, generalize more robustly across diverse environments and enhancement systems.

\subsection{Scatter plot analysis}
In this experiment, we visualize scatter plots of predicted versus ground-truth MOS scores to analyze how each method performs on the full test set. For the first case, NaiveEnsemble produces predictions that are compressed into a narrow mid-range of the MOS scale, resulting in moderate correlation and difficulty in capturing the full variability of human quality scores. MOS-SSL also suffers from a limited prediction range, with outputs clustered between 2.5 and 3.5, which weakens its ability to represent the full MOS spectrum and leads to lower correlation values. In contrast, GatherMOS-FS spreads predictions more broadly across the MOS scale, but the alignment with the ground-truth scores is weaker, as reflected by increased scatter around the diagonal trend. This suggests that although GatherMOS-FS improves score coverage, it produces less consistent alignment with ground-truth scores across the full range. Interestingly, GatherMOS-ZS, which integrates a zero-shot strategy and combines several acoustic features (RMS, ZCR, clipping ratio, and duration) with pseudo-labels, achieves better correlation and predictions that align more closely with the diagonal trend. Finally, GatherMOS-ZS*, which additionally incorporates MFCC and spectrogram features, produces the best alignment with ground-truth scores. This highlights that enriching the feature set with complementary acoustic information enables the model to generalize more reliably and allows the LLM to better calibrate predictions to reflect subjective human judgments across the full MOS range.

\graphicspath{ {./images/} }
\begin{figure}[t]
\centering
\includegraphics[width=8cm]{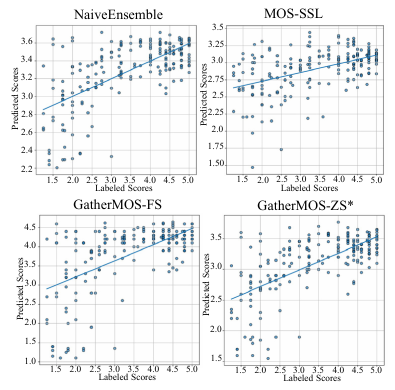} 
\caption{Scatter Plot Analysis Between GatherMOS and Other Models.} 
\label{fig:cross-domain}
\end{figure}

\section{Conclusion}
In this work, we explored the use of generative AI for non-intrusive speech quality estimation and introduced GatherMOS, a framework that acts as a meta-evaluator by combining acoustic feature representations with pseudo-labels from DNSMOS and VQScore. By leveraging the reasoning capabilities of large language models, GatherMOS integrates these diverse signals to produce more reliable MOS predictions. Our experiments show that incorporating richer acoustic features, such as MFCC and spectrogram statistics, consistently enhances performance. We also observed that few-shot in-context examples can provide substantial gains when the support set aligns with the test domain, but may reduce accuracy when mismatched. Across evaluations on the VoiceBank-DEMAND dataset, GatherMOS outperforms DNSMOS, VQScore, naive ensembles, and learning-based baselines trained on limited data. These findings suggest that LLM-based aggregation offers a promising direction for scalable and practical speech quality evaluation, reducing reliance on large annotated datasets while complementing existing approaches. Future work will extend the study to larger and more diverse datasets, and further explore strategies to improve the stability of few-shot prompting and the meta-evaluator.

\bibliographystyle{IEEEbib}
\bibliography{refs} 

\end{document}